# A novel framework based on deep learning and ANOVA feature selection method for diagnosis of COVID-19 cases from chest X-ray Images

Hamid Nasiri,[1] Seyyed Ali Alavi,[2]

[1] Department of Computer Engineering, Amirkabir University of Technology, Tehran, Iran.
[2] Electrical and Computer Engineering Department, Semnan University, Semnan, Iran.

Correspondence should be addressed to Hamid Nasiri; h.nasiri@aut.ac.ir

## Abstract

The new coronavirus (known as COVID-19) was first identified in Wuhan and quickly spread worldwide, wreaking havoc on the economy and people's everyday lives. Fever, cough, sore throat, headache, exhaustion, muscular aches, and difficulty breathing are all typical symptoms of COVID-19. A reliable detection technique is needed to identify affected individuals and care for them in the early stages of COVID-19 and reduce the virus's transmission. The most accessible method for COVID-19 identification is RT-PCR; however, due to its time commitment and false-negative results, alternative options must be sought. Indeed, compared to RT-PCR, chest CT scans and chest X-ray images provide superior results. Because of the scarcity and high cost of CT scan equipment, X-ray images are preferable for screening. In this paper, a pre-trained network, DenseNet169, was employed to extract features from X-ray images. Features were chosen by a feature selection method (ANOVA) to reduce computations and time complexity while overcoming the curse of dimensionality to improve predictive accuracy. Finally, selected features were classified by XGBoost. The ChestX-ray8 dataset, which was employed to train and evaluate the proposed method. This method reached 98.72% accuracy for two-class classification (COVID-19, healthy) and 92% accuracy for three-class classification (COVID-19, healthy, pneumonia).

## 1. Introduction

The new Coronavirus, also known as COVID-19, was initially discovered in Wuhan, China, in December 2019 [1]. COVID-19 is the name of the disease, and SARS-CoV-2 is the name of the virus. This novel infection spread from Wuhan to a large portion of China in less than 30 days [2]. Since November 19th, 2020, the COVID-19 pandemic had a detrimental effect on the world, with approximately 219,456,675 confirmed cases and 4,547,782 deaths reported till 27 September 2021; in addition, nearly 7.7 million workers have lost their jobs in America [3]. The majority of coronaviruses infect animals; however, due to their zoonotic nature, They have the ability to infect humans [4], and as a result, it has the potential to infect human airway cells, leading to pneumonia, severe respiratory infections, renal failure, and even death. Fever, cough, sore throat, headache, weariness, muscle soreness, and short breathing are common COVID-19 symptoms [5].

Vivid screening of infected individuals, aids them to be isolated and treated is a crucial and essential step in combating COVID-19 [1]. Reverse transcriptase-polymerase chain reaction (RT-PCR) testing, which can identify SARS-CoV-2 RNA from respiratory material, is the most common technique for detecting COVID-19 patients [6]. It requires specialized materials and equipment that are not readily available; because of the large number of false-negative results, it takes at least 12 hours, which is inconvenient considering that positive COVID-19 patients should be identified and followed up on as soon as possible [7], [8]. Chest-CT scan is another option for detecting the disease, which is more accurate than RT-PCR; For instance, 75% of negative RT-PCR samples had positive results on chest-CT scans [9]. CT scans have several drawbacks, including image collection time, related cost, and CT equipment availability [10]. When compared to CT scans, X-ray images are less expensive and more easily available [11]. As a result, the focus of the research is only on the use of X-ray imaging as a screening tool for COVID-19 patients.

Researchers discovered that COVID-19 patients' lungs contain visual markings such as ground-glass opacities—hazy darker areas that may distinguish COVID-19 infected individuals from non-infected patients [12], [13]. However, due to the limitations of experts, time constraints, and the irreversible consequences of misdiagnosis [4], it is crucial to discover a different approach to get faster and more reliable outcomes. The technological advancements facilitate the process of diagnosing the diseases, in other words, the widespread use of AI [14] mainly its areas such as machine learning and deep learning, are extremely constructive and researchers have made significant use of AI and deep learning in various medical areas [15]–[19]. CNN architecture is one of the most prominent deep learning techniques in the medical imaging field, with outstanding results [20].

Pre-trained neural networks are used in this paper, which is one of the most recent techniques. Using easily accessible pre-trained models, the proposed method extracts features from X-ray images. We utilize one of the feature selection methods in the second phase to acquire an appropriate number of features for classification. Finally, we use the XGBoost classifier to classify the specified features. The rest of the paper is organized as follows: Section 2 describes related works. In section 3, used materials and methods will be presented. In section 4, the experimental results are reported and analyzed. Finally, section 5 will present a summary of the findings and conclusions.

## 2. Related Works

Researchers worldwide are now trying to fight against the Covid-19; using radiological imaging and deep learning has made significant progress in this approach. Wang et al. [6] developed COVID-Net, a deep model for COVID19 detection that categorized normal, non-COVID pneumonia, and COVID-19 classes with 92.4 percent Accuracy. Apostolopoulos et al. [21] applied transfer learning and employed COVID-19, healthy, and pneumonia X-ray images to develop their model. Ozturk et al. [4] proposed using the DarkNet model to build a deep network. This model contains 17 convolution layers and utilizes the Leaky RelU activation function. For binary classes, this model was 98.08% accurate, and for multi-class cases, it was 87.02% accurate. Nasiri and Hasani [22] employed DenseNet169 to extract features from X-ray images and used XGBoost for classification; they gained 98.24% and 89.70% in binary, and multi-class classification, respectively. Sethy et al. [23] devised an in-depth feature combined support vector machine (SVM) based method for detecting coronavirus infected individuals using X-ray images. SVM is examined for COVID-19 identification utilizing the deep features of 13 different CNN models. Fareed Ahmad et al. [24] utilized X-ray images for training Deep CNN models like MobileNet, ResNet50, and InceptionV3 with a variety of options, including starting from scratch, fine-tuning with learned weights of all layers, and fine-tuning with learned weights and augmentation. Abbas et al. [25] verified a deep CNN termed Decompose, Transfer, and Compose (DeTraC) for COVID-19 chest X-ray image classification. Zhang et al. [26] presented the Parallel Channel Attention Feature Fusion Module (PCAF) and a new structure of convolutional neural network MCFF-Net based on PCAF. The network uses three classifiers to boost recognition efficiency: 1-FC, GAP-FC, and Conv1-GAP. Ucar and Korkmaz [27] developed the SqueezeNet that goes toward its light network design, is optimized for the COVID-19 detection with the Bayesian optimization additive.

Additionally Kang et al. [28] presented a transfer learning model that handles a dataset of COVID-19 infected patients' CT images. They achieved a test accuracy of 79.3%. Khan et al.

[1] represented CoroNet, a Deep Convolutional Neural Network model for diagnosing COVID-19 infection from chest X-ray images automatically. The proposed model is built on the Xception architecture. Narin et al. [29] proposed five models for diagnosing people with Pneumonia and Coronavirus using X-ray images.

Similarly He et al. [30] created a deep learning method to categorize COVID-19. They scanned 746 CT images, 349 of which were of infected patients and 397 of healthy people. The Self-Trans technique is proposed in this approach, which combines contrastive self-supervised learning with transfer learning to gain strong and unbiased feature representations while avoiding overfitting, resulting in a 94% accuracy rate. Xu et al. [31] applied deep learning techniques to create an early screening model to discriminate COVID-19 from influenza-A viral pneumonia and healthy cases using lung CT scans. Hemdan et al. [32] used 50 validated Chest X-ray images and 25 confirmed positive COVID-19 cases and developed the COVIDX-Net, which incorporates seven distinct architectures of deep convolutional neural network models, such as VGG19 as well as the second version of Google MobileNet. Minaee et al. [33] used publicly available datasets to build a dataset of 5000 chest X-rays. A board-certified radiologist discovered images that showed the existence of COVID-19 virus. Four prominent convolutional neural networks were trained to detect COVID-19 disease, using transfer learning

## 3. Materials and Methods

The proposed method employs the DenseNet169 deep neural network, as well as feature selection and the XGBoost algorithm, which will be discussed in the following section.

### 3.1. DenseNet169

A CNN's overall architecture is composed of two core parts: a feature extractor and a classifier. Convolution and pooling layers are the two essential layers of CNN architecture. Each node in the convolution layer extracts features from the input images by performing a convolution operation on the input nodes. Through averaging or calculating the maximum value of input nodes, the max-pooling layer abstracts the feature [34], [35]. DenseNet is a highly supervised network that contains a 5-layer dense block with a k = 4 rate of growth and the standard ResNet structure. Each layer's output in a DenseNet dense block includes the output of all previous layers, incorporating both low-level and high-level features of the input image, making it suitable for object detection [36]. The ILSVRC 2012 classification dataset, which was used for training DenseNet, contains 1,000 classes and 1.2 million images. The dataset images was cropped with the size of $224 \times 224$ before using as input for DenseNet. DenseNet presented a new connectivity pattern that introduced direct connections from any layer to all the following layers to improve information flow across layers even further [37]. In DenseNet, the *l* th layer, takes all feature maps $x_0, x_1, x_2, ..., x_{l-1}$ from the preceding layers as input, which is described by Equation (1).

$$x_l = H_l([x_0, x_1, x_2, ..., x_{l-1}]), \tag{1}$$

where $H_l(\cdot)$ is a singular tensor and $[x_0, x_1, x_2, ..., x_{l-1}]$ is the concatenated features from $l-1$ layers. To preserve the feature-map size constant, each side of the inputs is zero-padded by one pixel for convolutional layers with kernel size $3 \times 3$. DenseNet employed $1 \times 1$ convolution and $2 \times 2$ average pooling as transition layers between adjoining dense blocks. A global average pooling, in fact, is conducted at the end of the last dense block, and then a Softmax classifier is connected. In the three dense blocks, the feature-map sizes are $32 \times 32$, $16 \times 16$, and $8 \times 8$, respectively. On five distinct competitive benchmarks, this innovative architecture reached state-of-the-art accuracy for recognising the object [34], [37].

## 3.2. Analysis of Variance Feature Selection

New issues develop as a result of the creation of large datasets. Consequently, reliable and unique feature selection approaches are required [38]. Feature selection can assist with data visualization and understanding, as well as minimizing measurement and storage needs, training and utilization times, and overcoming the curse of dimensionality to enhance prediction performance [39], [40]. Analysis of variance (ANOVA) is a well-known statistical approach for comparing several independent means [41]. The ANOVA approach ranks features by calculating the ratio of variances between and within groups [42].

The ratio indicates how strongly the $\lambda$ th feature is linked to the group variables. Equation (2) is used to calculate the ratio $F$ value of $\lambda$ th g-gap dipeptide in two benchmark datasets:

$$F(\lambda) = \frac{s_B^2(\lambda)}{s_W^2(\lambda)} \tag{2}$$

where $s_B^2(\lambda)$ and $s_W^2(\lambda)$ are the sample variance between groups (also known as Mean Square Between, MSB) and within groups (also known as Mean Square Within, MSW), respectively, and can be calculated as Equation (3) and Equation (4).

$$s_B^2(\lambda) = \sum_{i=1}^{K} n_i \left( \frac{\sum_{j=1}^{n_i} f_{ij}(\lambda)}{n_i} - \frac{\sum_{i=1}^{K} \sum_{j=1}^{n_i} f_{ij}(\lambda)}{\sum_{i=1}^{K} n_i} \right)^2 / df_B \tag{3}$$

$$s_W^2(\lambda) = \sum_{i=1}^{K} \sum_{j=1}^{n_i} \left( f_{ij}(\lambda) - \frac{\sum_{i=1}^{K} \sum_{j=1}^{n_i} f_{ij}(\lambda)}{\sum_{i=1}^{K} n_i} \right)^2 / df_W \tag{4}$$

The degrees of freedom for MSB and MSW are $df_B = K - 1$ and $df_W = N - K$, respectively. The number of groups and total number of samples are represented by $K$ and $N$, respectively. The frequency of the $\lambda$ th feature in the $j$ th sample in the $i$ th group is denoted by $f_{ij}(\lambda)$. The number of samples in the $i$ th group is denoted by $n_i$ [43].

### 3.3. Extreme Gradient Boosting (XGBoost)

Chen and Guestrin proposed an efficient and scalable variation of the Gradient Boosting algorithm called Extreme Gradient Boosting (XGBoost). XGBoost has been widely employed by data scientists recently, and it had desirable results in a wide range of machine learning competitions [44], [45]. In certain ways, XGBoost differs from GBDT. First of all, the GBDT algorithm only employs a first-order Taylor expansion, whereas XGBoost augments the loss function with a second-order Taylor expansion. Secondly, the objective function uses normalization to prevent overfitting and reduce the method's complexity [46], [47]. Third, XGBoost is extremely adaptable, allowing users to create their own optimization objectives and evaluation criteria. Nevertheless, by establishing class weight and using AUC as an assessment criterion, the XGBoost classifier can handle unbalanced training data efficiently. In summary, XGBoost is a scalable and flexible tree structure improvement model that can manage sparse data, enhance algorithm speed, and minimize computing time and memory for large-scale data [48].

Formally, the XGBoost algorithm can be described as follows:

Given a training dataset of $n$ samples $T = \{(\mathbf{x}_1, y_1), (\mathbf{x}_2, y_2), ..., (\mathbf{x}_n, y_n)\}, \mathbf{x}_i \in \mathbb{R}^m, y_i \in \mathbb{R}$, the objective function can be defined by:

$$obj(\theta) = \sum_{i}^{n} l(y_i, \hat{y}_i) + \sum_{t=1}^{T} \Omega(f_t) \tag{5}$$

where $l(y_i, \hat{y}_i)$ measures the difference between the target $y_i$ and the prediction $\hat{y}_i$ and $f_t$ denotes the prediction score of $t$ th tree. The estimated loss function can be computed based on Taylor expansion of the objective function:

$$L^{(t)} \simeq \sum_{i=1}^{k}\left[l(y_i, \hat{y}^{(t-1)}) + g_i f_t(x_i) + \frac{1}{2} h_i f_t^2(x_i)\right] + \Omega(f_t) \qquad (6)$$

where $g_i = \partial_{\hat{y}^{(t-1)}} l(y_i, \hat{y}^{(t-1)})$ denotes each sample's first derivative and $h_i = \partial^2_{\hat{y}^{(t-1)}} l(y_i, \hat{y}^{(t-1)})$ denotes each sample's second derivative, and the first and second derivatives of each data element are all that the loss function requires [49].

### 3.4. Proposed Method

In this study, pre-processing methods were employed on the dataset, which includes label encoder for classes and using normalization on images and as a result, less redundant data are given as the input to the network. Deeply influenced by the brain's structure, deep learning as a sub-field of machine learning was emerged. In the area of medical image processing, as in many other areas, deep learning approaches have continued to demonstrate excellent results in past years [29]. ImageNet is a dataset of millions of images organized into 1000 categories when it comes to image processing. The next step was to apply several pre-trained models that were trained based on this dataset. Densnet169 had the best performance among those models, so it was selected as the feature extractor in the proposed method. The X-ray dataset images scaled at a fixed size of $224 \times 224$ pixels, which is the DenseNet169 input size.

The final layer of the DenseNet169 network, which was used to predict ImageNet dataset labels, was eliminated. Global average pooling, a pooling method designed to substitute fully connected layers in classical CNNs, was added in the final layer of the network. One of the benefits of global average pooling is that there are no parameters to adjust in this layer; therefore, no training is needed. Additionally, because global average pooling sums up the dimensional information, it is more resistant to input dimensional translations [50]. The X-ray images were given to the network in order to extract features from DenseNet169, and 1664 features were extracted as a result.

When a learning model is given many features and few samples, it is likely to overfit, causing its performance to degrade. Among researchers, feature selection is a widely used strategy for reducing dimensionality [51]. In order to reduce the classification time and increase the classifier performance, the ANOVA feature selection method was employed to reduce the number of features. Thus, the range of 50 to 500 features was applied for the purpose of selecting the best number of features for classification (using validation set). Finally, the selected features were given to the XGBoost for detection of COVID-19. Figure 1 shows the general framework of the proposed method.

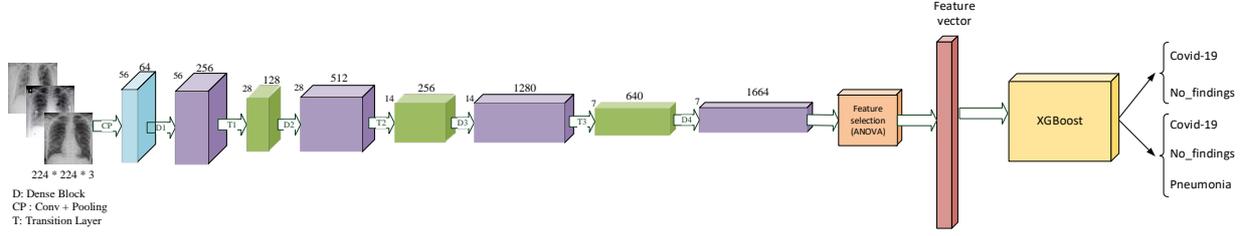

Figure 1: The architecture of proposed model.

## 4. Results and Discussion

Several performance metrics such as precision, recall, specificity, and $F_1$-Score, as well as accuracy were utilized, to evaluate several deep learning models with the proposed methodology, because accuracy alone cannot evaluate a model's usefulness [52]. Accuracy is the ratio of number of correctly predicted samples to the total number of samples. The Equation (7) can be used to calculate accuracy.

$$Accuracy = \frac{TP + TN}{Total} \tag{7}$$

Precision is the proportion of predicted true positive values to the total number of predicted true positive and false positive values. A model with a low precision is prone to a high false-positive rate. Precision can be calculated using Equation (8).

$$Precision = \frac{TP}{FP + TP} \tag{8}$$

The number of true positives divided by the sum of true positives and false negatives is known as recall or sensitivity. When there is a large cost associated with false negatives, the model statistic used to pick the optimal model is recall. Recall can be computed using Equation (9).

$$Recall = \frac{TP}{TP + FN} \tag{9}$$

Specificity is the proportion of predicted true negatives to the summation of predicted true negatives and false positives. Specificity can be determined using Equation (10).

$$Specificity = \frac{TN}{TN + FP} \quad (10)$$

$F_1$-Score combines precision and recall. As a result, both false positives and false negatives are included while calculating this score. It is not obviously as simple as accuracy for comparison. However, $F_1$-Score is generally more valuable than accuracy, particularly if the problem is an imbalanced classification problem. Equation (11) can be used to calculate the $F_1$-Score.

$$F_1\text{-}Score = \frac{2 \times Recall \times Precision}{Recall + Precision} \quad (11)$$

In this study, the dataset that Ozturk et al. [4] collected has been employed, gathered from two distinct sources, and includes COVID-19, No-findings, and Pneumonia as shown in Figure 2. This first class of dataset contained 43 women, and 82 men confirmed they were infected with COVID-19. The average age of 26 COVID-19 confirmed individuals is about 55 years old, according to the age info supplied. The remaining two classes were chosen randomly from the Wang et al. [53] ChestX-ray8 dataset, which included 500 No-findings and 500 Pneumonia images.

Two distinct perspectives were conducted to identify and classify COVID-19. First, the proposed technique was validated in order to classify binary classes labelled COVID-19 and No-findings. Second, the proposed approach was used to classify three different groups: COVID-19, No-findings, and Pneumonia. In the first aspect, the two-class problem, the proposed method effectiveness is measured using the 5-fold cross-validation. A total of 80% of the dataset was used for training and 20% for testing. Following the extraction of features by Densenet-169, ANOVA selected 67 features from 1664 as an optimal number for classification, as a result, about 96% of features are reduced, and the XGBoost classification process was significantly sped up.

The 5-fold cross-validation had an average accuracy of 98.72%, and the confusion matrix was computed for each fold and overlapped, as shown in Figure 3. The total of the confusion matrix entries acquired in all folds is used to generate the overlapping confusion matrix. As a consequence, the goal is to get a sense of the model's general patterns [4]. It shows that the proposed architecture correctly identified COVID-19 and No-findings with 100% and 98.43% accuracy, respectively. In other words, the proposed method performs better at detecting true-positive samples.

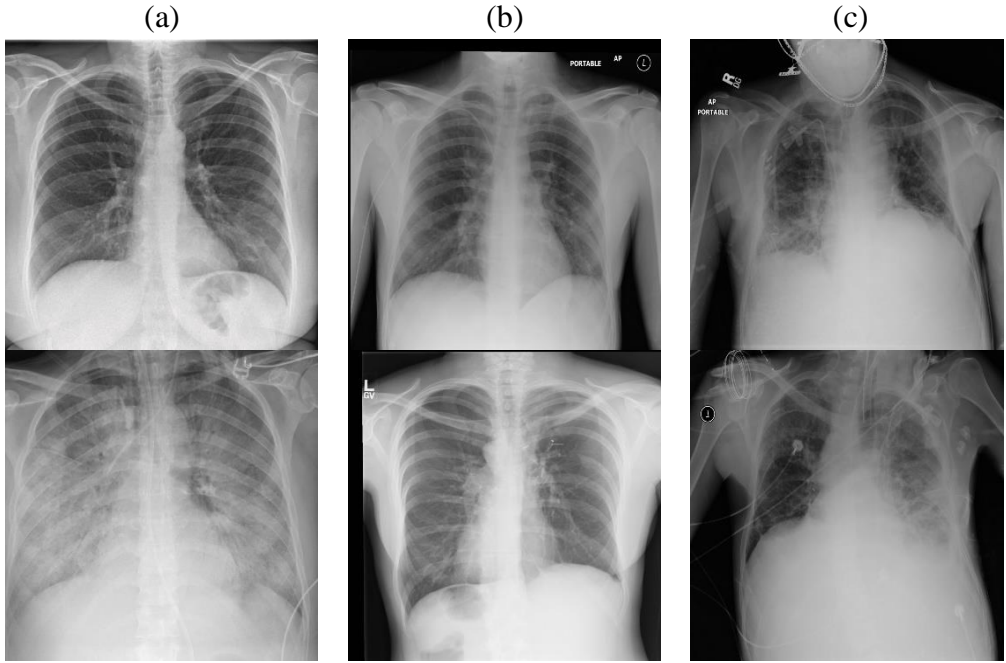

Figure 2: Representation of lungs X-ray in COVID-19 patients (a), No-findings (b), and patients with Pneumonia (c).

Precision, recall, specificity, and $F_1$-Score values achieved are 99.21%, 93.35%, 100%, and 97.87%, respectively. Table 1 represents the comparison of the proposed method with Ozturk et al. [4] and Nasiri and Hasani [22] in terms of accuracy, precision, recall, specificity, and $F_1$-Score values for each fold and the average of all folds, which Nasiri and Hasani [22] had better results than Ozturk et al. [4] and the proposed method outperforms them all except recall.

In the multi-class problem, 80% of X-ray images dataset was used for training and the 20% remaining employed for evaluation of proposed architecture. ANOVA was used to select 275 features out of 1664 as the ideal number for classification. As a consequence, almost 84% of features are decreased, and XGBoost classification process was substantially ramped up and performance improved. The accuracy of validation set achieved 92% and the confusion matrix was illustrated as Figure 4. Like binary class problem, this confusion matrix indicates that the proposed method had a stronger result in finding COVID-19 rather than No-findings, and Pneumonia. Precision, recall, specificity, and $F_1$-Score values of 94.07 %, 88.46%, 100%, and 92.42%, were reached respectively. In terms of accuracy, precision, recall, specificity, and $F_1$-Score values of the validation set, Table 2 compares the proposed approach to Ozturk et al. [4] and Nasiri and Hasani [22].

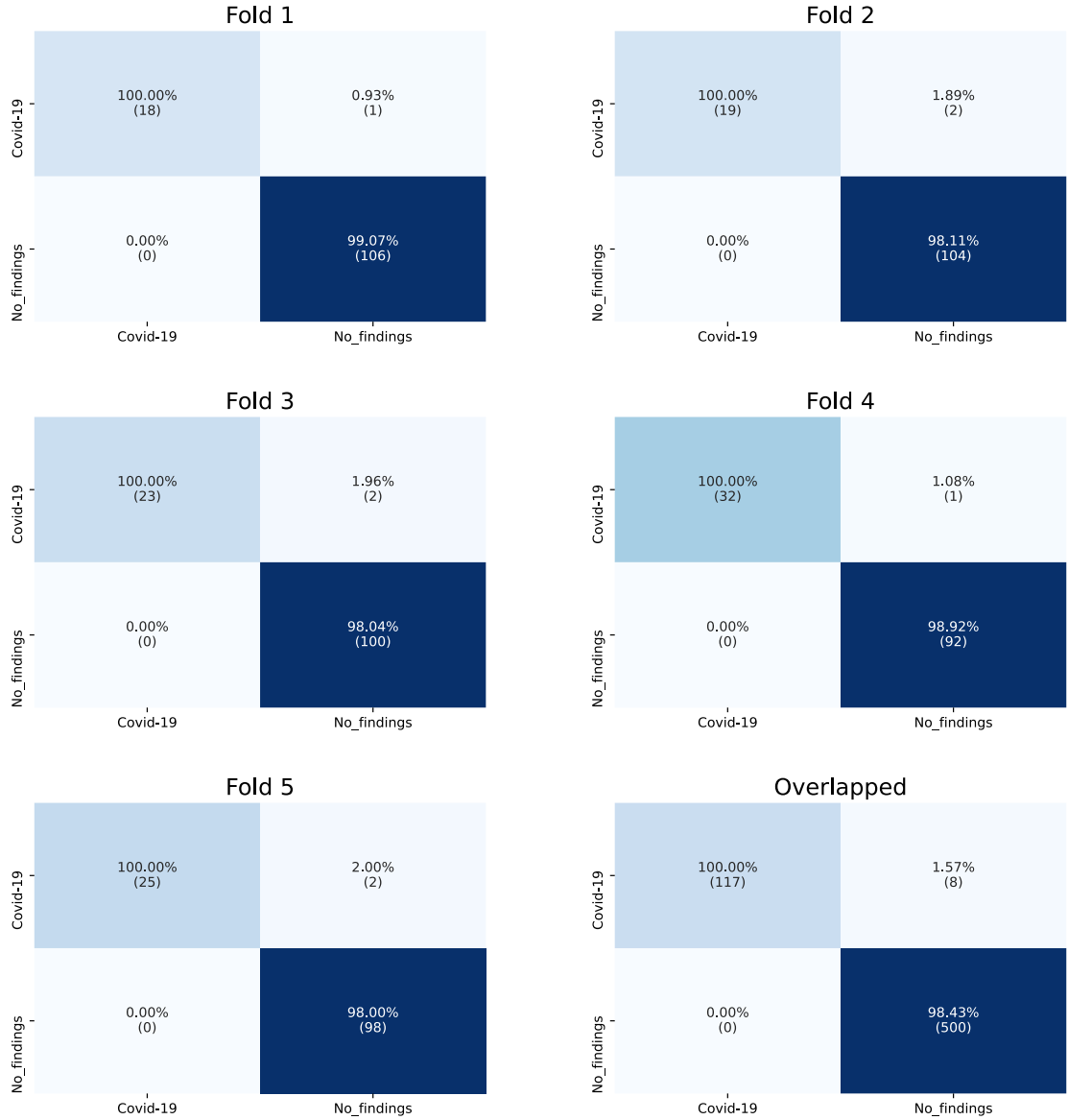

Figure 3: Confusion matrix for binary class problem.

Table 1 Comparison of the proposed method with other methods in binary class problem

| Performance metrics (%) | Methods | Folds-1 | Folds-2 | Folds-3 | Folds-4 | Folds-5 | Average |
|---|---|---|---|---|---|---|---|
| Recall | Proposed method | 94.73 | 90.47 | 92.00 | **96.96** | 92.59 | 93.33 |
|  | Ozturk et al. | **100** | **96.42** | 90.47 | 93.75 | **93.18** | **95.13** |
|  | Nasiri and Hasani | 95.20 | 95.40 | **96.70** | 81.40 | 91.40 | 92.08 |
| Specificity | Proposed method | 100 | 100 | 100 | **100** | 100 | **100** |
|  | Ozturk et al. | 100 | 96.42 | 90.47 | 93.75 | 93.18 | 95.30 |
|  | Nasiri and Hasani | 100 | 100 | 100 | 89.90 | 100 | 99.78 |
| Precision | Proposed method | 99.53 | 99.05 | 99.01 | **99.46** | 99.00 | **99.21** |
|  | Ozturk et al. | **100** | 94.52 | 98.14 | 98.57 | 98.58 | 98.03 |
|  | Nasiri and Hasani | 99.50 | **99.50** | **99.40** | 95.30 | **99.02** | 98.54 |
| $F_1$-Score | Proposed method | 98.41 | 97.02 | 97.42 | **98.96** | **97.57** | **97.87** |
|  | Ozturk et al. | **100** | 95.52 | 93.79 | 95.93 | 95.62 | 96.51 |
|  | Nasiri and Hasani | 98.50 | **98.50** | **98.20** | 92.50 | 97.30 | 97.00 |
| Accuracy | Proposed method | 99.20 | 98.40 | 98.40 | **99.20** | 98.40 | **98.72** |
|  | Ozturk et al. | **100** | 97.60 | 96.80 | 97.60 | 97.60 | 98.08 |
|  | Nasiri and Hasani | 99.20 | **99.20** | **99.20** | 95.20 | 98.40 | 98.24 |

The proposed method applied on ten pre-trained networks for both binary and multi-class problem. As shown in Table 3, The average of 5-fold cross-validation accuracy employed to compare approaches in two class problem whereas the best fold accuracy was used to compare approaches on multi-class problem. DenseNet169 outperforms other pre-trained networks in both binary and multi-class problem. Additionally, the gradient-based class activation mapping (Grad-CAM) [54] was used to represent the decision area on a heatmap. Figure 4 illustrates the heatmaps for three COVID-19 cases, confirming that the proposed method extracted correct features for detection of COVID-19, and the model is mostly concentrated on the lung area. Radiologists might use these heatmaps to evaluate the chest area more accurately.

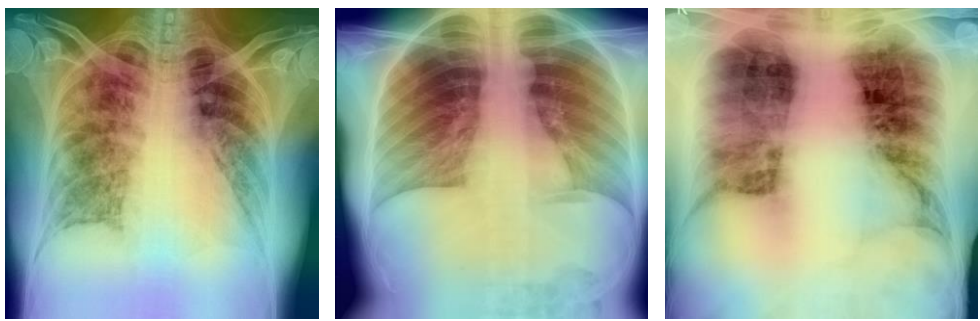

Figure 4: The heatmap of three confirmed COVID-19 X-ray images.

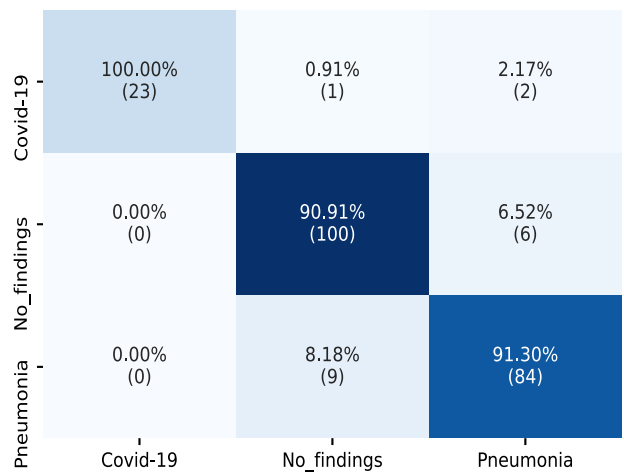

Figure 5: Confusion matrix for three-class problem.

Table 2 Comparison of the proposed method with other methods in multi-class problem

| | Performance metrics (%) | | | | |
|---|---|---|---|---|---|
| Methods | Recall | Specificity | Precision | $F_1$-score | Accuracy |
| Proposed method | 88.46 | **100** | **94.07** | **92.42** | **92.00** |
| Ozturk et al. | 88.17 | 93.66 | 90.97 | 89.44 | 89.33 |
| Nasiri and Hasani | **95.20** | **100** | 92.50 | 91.20 | 89.70 |

Table 3 Comparison of different deep neural networks

| DNN | Binary Class Accuracy | Multi Class Accuracy |
| --- | --- | --- |
| DenseNet169 | **98.72%** | **92%** |
| InceptionV3 | 92.96% | 82.22% |
| NASNetLarge | 94.88% | 84.00% |
| ResNet152 | 94.71% | 77.33% |
| VGG16 | 97.43% | 88.88% |
| VGG19 | 97.28% | 88.88% |
| Xception | 95.68% | 80.88% |
| EfficientNetB0 | 97.92% | 88.88% |
| InceptionResNetV2 | 94.88% | 83.11% |

The proposed method was compared to relevant works in Table 4. Wang et al. [6] applied 16,756 X-ray chest images from diverse sources to develop COVID-Net and achieved a 92.4% accuracy rate on 3-class classification problem. Sethy et al. [23] utilized ResNet150 to extract features from 25 COVID-19 positives and 25 healthy chest X-ray images, then used SVM to classify them, achieving a 95.38 percent accuracy rate. Wang et al. [26] proposed M-Inception for 195 COVID-19 infected patients and 258 regular cases, and as a result, 92.8% accuracy rate was achieved. Hemdan et al. [32] trained and evaluated COVIDX-Net using 25 confirmed COVID-19 and 25 non-infected cases X-ray images, achieving a 90% Accuracy rate.

Narin et al. [29] used 50 public source COVID-19 chest X-ray images and 50 normal images from another source to test three alternative CNN models, obtaining a 98% accuracy rate. Sethy et al. [23] reached 95.38% accuracy using ResNet50 and SVM which was evaluated by 50 X-ray images. Ying et al. [55] employed 777 confirmed COVID-19 patients and 708 normal cases CT images to develop their deep model base on the pre-trained network model, ResNet50, which reached 86% accuracy rate. Apostolopoulos et al. [21] scored 98.75% and 93.84% success for binary and multi-class classification problem, respectively. They used 224 confirmed COVID-19, 700 pneumonia, and 504 normal X-ray images to evaluate their model based on the pre-trained model, VGG-16, and transfer learning.

Zheng et al. [56] gained 90.8% accuracy employing CT images of 313 positive COVID-19 and 229 normal cases to develop their model. Xu et al. [31] applied ResNet on the dataset of 219 confirmed COVID-19 and 224 pneumonia and 175 normal CT images, scoring 86.7% performance. Ozturk et al. [4] used 125 positive COVID-19, 500 No-findings, and 500 pneumonia X-ray images to develop their model, resulting in 98.08% for binary-class and 87.02% multi-class accuracy rate. The dataset that Ozturk et al. [4] gathered from various sources has been used in this paper. For two-class and multi-class classification problem, 92% and 98.72% accuracy are obtained, respectively, in this paper. Table 4 shows that the proposed approach outperforms most of the existing deep learning-based models in terms of accuracy. However, it should be emphasized that the findings in Table 4 were derived from different datasets and different experimental setups.

Table 4 Comparison of the proposed method with other DNN based methods

| Study | Type of Images | Number of Cases | Method Used | Accuracy (%) |
|---|---|---|---|---|
| Apostolopoulos et al. [21] | Chest X-ray | 224 COVID-19 (+)<br>700 Pneumonia<br>504 Healthy | VGG-19 | 93.48 |
| Wang et al. [6] | Chest X-ray | 53 COVID-19 (+)<br>5526 COVID-19 (−) | COVID-Net | 92.4 |
| Sethy et al. [23] | Chest X-ray | 25 COVID-19 (+)<br>25 COVID-19 (−) | ResNet50+SVM | 95.38 |
| Hemdan et al. [32] | Chest X-ray | 25 COVID-19 (+) | COVIDX-Net | 90.0 |
| Narin et al. [29] | Chest X-ray | 50 COVID-19 (+)<br>50 COVID-19 (−) | Deep CNN ResNet50 | 98 |
| Ying et al. [55] | Chest CT | 777 COVID-19 (+)<br>708 Healthy | DRE-Net | 86 |
| Wang et al. [26] | Chest CT | 195 COVID-19 (+)<br>258 COVID-19 (−) | M-Inception | 82.9 |
| Zheng et al. [56] | Chest CT | 313 COVID-19 (+)<br>229 COVID-19 (−) | UNet+3D Deep Network | 90.8 |
| Xu et al. [31] | Chest CT | 219 COVID-19 (+)<br>224 Viral Pneumonia<br>175 Healthy | ResNet +Location Attention | 86.7 |
| Ozturk et al. [4] | Chest X-ray | 125 COVID-19 (+)<br>500 No-Findings | DarkCovidNet | 98.08 |
| | | 125 COVID-19 (+)<br>500 No-Findings<br>500 Pneumonia | | 87.02 |
| **Proposed Method** | Chest X-ray | 125 COVID-19 (+)<br>500 No-Findings | DenseNet169+ ANOVA+XGBoost | **98.72** |
| | | 125 COVID-19 (+)<br>500 No-Findings<br>500 Pneumonia | | **92.0** |

## 5. Conclusion

Early diagnosing COVID-19 is a crucial step to prevent mortality and transmission of the virus. RT-PCR is the most accessible tool to identify COVID-19, but finding other alternatives for this problem is essential due to false-negative outcomes and time limitations. Chest CT and X-ray images are suitable substitutes for RT-PCR, but because of the lack of CT hardware, X-ray images are a superior tool for diagnosing COVID-19. AI and machine learning technologies play a virtual role in the quicker detection of COVID-19. In this study, a pre-trained model, DenseNet169, was utilized to extract features from X-ray images, and ANOVA was employed to select features to decrease classification time and improve performance. And finally, selected features were classified using XGBoost. The ChestX-ray8 dataset was used to evaluate the proposed method. Experimental results show that the proposed method obtained 98.72% and 92% accuracy in two and three-class problems, respectively, and outperforms other state-of-the-art methods.

## Conflicts of Interest

The authors declare that they have no conflicts of interest.

## Funding Statement

This research received no specific grant from any funding agency in the public, commercial, or not-for-profit sectors.

## Data Availability

Publicly available ChestX-ray8 dataset was used in this study, which is available at https://github.com/seyyedalialavi2000/COVID-19-detection

## Code Availability

The source code of the proposed method required to reproduce the predictions and results is available at https://github.com/seyyedalialavi2000/COVID-19-detection